\documentclass[final]{pasj00}

\Received{}
\Accepted{}
\Published{}
\SetRunningHead{Astronomical Society of Japan}{Usage of \texttt{pasj00.cls}}

\usepackage{times}

\begin{document}

\title{What Kinds of Accretion Disks Are There in the Nuclei of Radio Galaxies?}

\author{Osamu \textsc{Kaburaki},\thanks{Present address: Otomaru-machi 249, 
Hakusan-shi, Ishikawa 924-0826} 
Takanobu \textsc{Nankou}, 
Naoya \textsc{Tamura}, and 
Kiyoaki \textsc {Wajima}}
\affil{Domain for Fundamental Sciences, Graduate School of Science and Engineering, \\
Yamaguchi University, Yamaguchi 753--8512}
\email{okab@amail.plala.or.jp, wajima@yamaguchi-u.ac.jp}

\KeyWords{galaxies: accretion, accretion disks --- galaxies: jets --- galaxies: individual 
 (NGC 4261) --- magnetic field}

\maketitle

\begin{abstract}

It seems to be a widely accepted opinion that the types of accretion disks (or flows) generally 
realized in the nuclei of radio galaxies and in further lower mass-accretion rate nuclei are inner, 
hot, optically thin, radiatively inefficient accretion flows (RIAFs) surrounded by outer, cool, 
optically thick, standard type accretion disks.  
However, observational evidence for the existence of such outer cool disks in these nuclei is 
rather poor. Instead, recent observations sometimes suggest the existence of inner cool disks 
of non-standard type, which develop in the region very close to their central black holes. 
Taking NGC 4261 as a typical example of such light eating nuclei, for which both flux data ranging 
from radio to X-ray and data for the counterjet occultation are available, we examine the plausibility 
of such a picture for the accretion states as mentioned above, based on model predictions. 
It is shown that the explanation of the gap seen in the counterjet emission in terms of the 
free-free absorption by an outer standard disk is unrealistic, and moreover, the existence itself 
of such an outer standard disk seems very implausible. 
Instead, the model of RIAF in an ordered magnetic field (so called resistive RIAF model) can well 
serve to explain the emission gap in terms of the synchrotron absorption, as well as to reproduce 
the observed features of the overall spectral energy distribution (SED). 
This model also predicts that the RIAF state starts directly from an interstellar hot gas phase 
at around the Bondi radius and terminates at the inner edge whose radius is about 100 times the 
Schwartzschild radii. 
Therefore, there is a good possibility for a cool disk to develop within this innermost region.  

\end{abstract}

\section{Introduction}

Now, it has become a great paradigm in astrophysics that accretion flows onto compact objects 
drives various types of activities, such as those seen in the active galactic nuclei (AGNs) 
and their associated jets, $\gamma$-ray bursters, bursts and jets in Galactic X-ray binaries 
and so on. However, the state of affairs is still far from settled as for the detailed specifications 
of the nature (or kind) of related accretion disks (or flows) in these various circumstances. 

It is commonly accepted that the most essential parameter to control the states of accretion flows 
is the dimension-less mass accretion rate, $\dot{m}$, normalized to the Eddington rate of a central 
object (\cite{FKR02, Arm04, KFM08}). 
When $\dot{m}\sim 1$, geometrically thin, radiatively efficient so-called standard disks \citep{SS73} 
would develop, and when $\dot{m}$ further exceeds a certain limit, optically thick but radiatively 
inefficient accretion flows (often called slim disks) appear. Therefore, both these states are expected 
to be realized in luminous AGNs such as quasars and Seyfert galaxies whose accretion rates are believed 
to be large and amount to a considerable fractions of the Eddington limit. 
 
On the other hand, when the accretion rate is very small in the sense that $\dot{m}\ll 1$, the flows 
become optically thin and radiatively inefficient. Usually, only this specific branch is called 
the radiatively inefficient accretion flows (RIAFs). RIAFs are, therefore, expected to appear 
in the low luminosity AGNs (LLAGNs) such as the nuclei of low-excitation radio galaxies (LERGs, 
see e.g., \cite{EHC07}) and of most nearby galaxies (\cite{Ho08}), at least in the inner parts 
of their accretion regions (e.g., \cite{Nry02}). 

The explicit doubts for the presence of pure standard disks in luminous AGNs have been cast, e.g., 
by \citet{Lhy99} and \citet{Gsk08}. Leahy insists that the origin of the big blue bump (BBB) needs 
reconsideration because i) the lack of strong EUV lines suggests the emission not caused by 
bremsstrahlung, ii) the accretion disk cannot be a simple standard one since the observed spectra 
extend to X-rays and iii) the disk should be much smaller than that conceived in the standard model, 
as the tight correlation within 1 hour of the variations in UV and optical bands suggests. The doubt 
of Gaskell is mainly based on the facts that i) the overall shapes of the spectral energy distribution 
(SED) of AGNs are nearly flat excluding BBB in the sense that $\nu F_{\nu}\sim$ const., where $\nu$ 
and $F_{\nu}$ are the frequency and associated flux, respectively, and ii) the frequency 
dependence of $F_{\nu}$ in BBB largely deviates from the prediction of the standard disk model, 
$F_{\nu}\propto\nu^{1/3}$. 

For LLAGNs, the absence of the standard disk, at least in its original form, has been suspected 
observationally. It is reported (e.g., \cite{CCC99,CCC00}) that in most of the FR I samples examined 
by {\em Hubble Space Telescope} ({\em HST}) there exist unresolved optical sources ($\lesssim 0.01$ pc), 
named central compact cores (CCCs), which show good correlation with the flux of radio cores. 
However, the evidence for the presence of standard disks is not detected, and the disks are supposed 
to be of RIAF type. 

Recently, \citet{Mz07} argues that in SED of most LLAGNs ($\sim10^{40}$ erg s$^{-1}$) the 
contribution from UV is not weak and is similar to those of Seyfert 1 ($\sim10^{44}$ erg s$^{-1}$), 
though the radio/UV ratio increases according to the decreasing luminosity of LLAGNs. This means that 
the emission ranging from UV to X-ray is rather insensitive to AGN luminosity, whereas the radio 
luminosity increases as AGN luminosity decreases. He insists, therefore, that there exist some type 
of cool thermal disks with low accretion rates in the inner parts of most nearby LLAGNs. The existence 
of inner cool disks in LLAGNs has also been put forward based on the analogy between the low/hard 
state of Galactic X-ray binaries and the state in jet-dominated LLAGNs (e.g., \cite{LPK03}). In the 
case of X-ray binaries, many observational evidences for the presence of inner cool disks are now 
accumulating (e.g., \cite{MHM06, Ryk07}). 

For the class of AGNs in which the mass accretion rates are believed to be larger than in LLAGNs, 
the presence of the inner cool disks is suggested by observations even more strongly. For example, 
in broad-line radio galaxies (BLRGs), which are classified as type 1 AGNs (see, e.g., \cite{Ury95}) 
and belong to the class of high-excitation radio galaxies (HERGs, see e.g., \cite{EHC07}) that have 
larger accretion rates than LERGs, the existence of untruncated cool, thermal disks almost 
down to the inner-most stable circular orbit (ISCO) is insisted (e.g., \cite{BRF02, BF05}). 
Their spectral fittings further suggest that the cool disks do not need to extend beyond several 
or tens of the Schwarzschild radii. For Seyfert galaxies, the presence of a compact cool disk 
around each central black hole has been established more firmly from X-ray spectra obtained by 
{\em XMM-Newton} (e.g., \cite{Nan06}).  

It is interesting to see that the importance of inner cool disks has begun to draw much attention 
also from theoretical side. In fact, the presence of such geometrically-thin, optically-thick 
accretion disks in the innermost regions of accretion flows have been proposed for a long time 
(e.g., \cite{PRP73, RBBF82, BSR87, HM91, SZ94, FR95, MF02, LPK03}; \authorcite{KnB04}
\yearcite{KnB04, KnB07}). Among them, \citet{KnB07} clearly demonstrate the role played by the 
vertical component of a global magnetic field in extracting energy and angular momentum from 
the accretion disk and in dissipating energy in the corona sandwiching the disk. Such a disk-corona 
structure developed very close to the central black hole is now believed to be the most plausible 
site of jet launching. 

It is often argued that a RIAF develops only within the truncation radius $r_{t}$, which forms the 
inner edge of a standard disk extending outside (\cite{EMN97, NMQ98, Cao04, Yun07}). 
Observationally, however, the nucleus of our Galaxy (Sgr A$^*$) has been marginally resoled by 
{\em Chandra} X-ray Observatory revealing that the hot accretion flow starts roughly from its Bondi 
radius (\cite{Bag03}, see also the spectral fittings by \cite{YKK02}). Also, there is a proposition 
based on observations of radio-loud AGNs that there are two phases for their accretion flows, in which 
LERGs accrete hot surrounding gas directly from their Bondi radii whereas HERGs accrete 
cold dense gas (\cite{Bst05}, \cite{HEC07}, \cite{EHC07}). 

Standing in such a new trend as reviewed above, we concentrate in this paper especially on 
the problem, if or not a standard-type accretion disk can exist in the outer region of 
an accretion flow of radio galaxies and LLAGNs. 
As a typical example of such low accretion-rate nuclei, we discuss the case of NGC 4261. However, 
we expect that the results obtained here would apply generally to the galaxies of similar type.
The main clue to solve our problem is the occultation of counterjet by its associated accretion 
disk. 

In \S 2, the observational facts related to the emission gap in the counterjet of NGC 4261 are described 
briefly. The implausibility of interpreting this gap by the free-free absorption of a radiatively-cooled 
accretion disk of conceivable types is discussed in \S 3, and the plausibility of interpreting it as 
the synchrotron absorption of a RIAF in a global magnetic field is roughly shown in \S 4. 
The reliability of the latter interpretation is further confirmed in \S 5 by performing 
numerical calculations based on our own RIAF model, in the course of global spectral fittings.
The final section summarizes the results obtained. 

\section{Counterjet Absorption in NGC 4261}

NGC 4261 is an FR I radio galaxy located at 30 Mpc from our Galaxy and the mass of its central black 
hole is estimated to be $(5\pm 1)\times 10^8 M_{\odot}$ (\cite{FFJ96}). As discussed in detail in 
\authorcite{Jon00} (\yearcite{Jon00, Jon01}), the emission from bipolar radio jets shows a narrow gap 
on the east side of the central core, which is regarded as the counterjet. They have pointed out that 
this gap can be interpreted as an absorption caused by the accretion disk surrounding the central 
black hole. 

\begin{figure}
 \begin{center}
  \FigureFile(85mm, 80mm){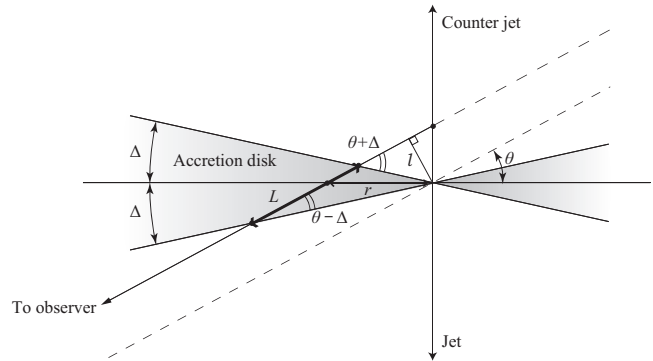}
 \end{center}
 \caption{An illustration of the geometrical configuration considered in the text. The polar axis 
indicates the direction of jet and counterjet. Suspected presence of an inner cool disk is neglected 
here.}
\label{fig:Figure1}
\end{figure}

Here, we quote from their papers the values of the geometrical parameters (see Fig.\ \ref{fig:Figure1}) 
that are necessary for the following discussion. The angle between the equatorial plane (assumed to be 
normal to the jet axis) and a line of sight is $\theta=26^{\circ}$. The half-opening angle of the 
geometrically thin accretion disk, $\Delta=\phi/2= 6.5^{\circ}$ (full-opening angle is $\phi$ according 
to their notation), is inferred from the HI observations of \citet{vLan00}. 
Therefore, the projected distance from the center to the line of sight, 
\begin{equation}
  l = r\sin\theta = r^*\sin(\theta-\Delta), 
\end{equation} 
becomes $l=0.44\ r$, where $r$ is the equatorial distance and $r^*$ is the surface distance from 
the center, respectively. Further, the path length through the disk, 
\begin{equation}
  L = l\{\cot(\theta-\Delta) - \cot(\theta + \Delta)\}, \quad (\theta\geq\Delta>0)
\end{equation}
is $L=1.3\ l$. 

Combining these results, we have $L(x)= 0.57 xr_{\rm S}$ for the line of sight which crosses the 
equatorial plane at a given radius $r$, where $x\equiv r/r_{\rm S}$ is the non-dimensional radius 
normalized to the Schwarzschild radius, $r_{\rm S}=2GM/c^2$, of the central black hole of mass $M$. 
\citet{Jon01} have derived from their observation at 8.4 GHz that $l=0.23$ pc for the projected 
distance from the core to the center of the coujerjet absorption. This value corresponds to 
$r=0.52$ [pc] ($x=1.0\times10^4$) and $L=0.3$ [pc], respectively. 

When the absorption coefficient $\alpha_{\nu}(x)$ at a given frequency $\nu$ is given on the 
equatorial plane as a function radius $r$, we estimate the optical thickness across the disk 
along the line of sight roughly as 
\begin{equation}
  \tau_{\nu}(x) = \frac{1}{2}\alpha_{\nu}(x)L(x). \label{eq:tau}
\end{equation} 

\section{Free-free Absorption by Radiatively Cooled Disks}

Although \authorcite{Jon00}(\yearcite{Jon00, Jon01}) discussed the possibility of interpreting 
the gap in the counterjet as due to the free-free absorption by an optically thin accretion disk 
developed in its nuclear region, it does not seem to be plausible. This is because they fix the 
temperature in the disk arbitrarily at $\sim 10^4$ K and hence the consistency of their model is 
not guaranteed. In fact, if the disk is optically thin and hence radiatively inefficient, the gas 
temperature is expected to rise close to the virial one. Then, the dominant cooling mechanism is not 
the free-free emission but would be the inverse Compton cooling. The self-consistent model for 
such a situation is known as SLE model (\cite{SLE76, KFM08}), which is however thermally unstable. 
Thus, the absorption of counterjet emission by a radiation-cooled, optically-thin disk is anyway 
implausible.

Therefore, we examine next the possibility that the absorption is due to a radiatively-efficient, 
optically-thick disk of the standard type. According to the results of \citet{SS73}, we cite the 
expressions for relevant physical quantities in the standard accretion disk. They are expressed 
as functions of the normalized radius $x$ and subsidiary three parameters. The parameters are all 
non-dimensional quantities. They are the viscosity parameter $\alpha$, the mass of a central black 
hole $m\equiv M/10^8M_{\odot}$, and the mass accretion rate $\dot{m}\equiv \dot{M}/\dot{M}_{\rm Edd}$ 
in the disk, where $M_{\rm Edd}=L_{\rm Edd}/c^2$ is the Eddington accretion rate. Note that the 
normalizations adopted here for $x$, $m$ and $\dot{m}$ are different in part from both of those 
adopted in \citet{FKR02} and \citet{KFM08}. 

Generally, the standard disk is divided into three regions according to the physical conditions 
realized in it. The opacity is dominated by the electron scattering in the inner and middle regions, 
whereas the free-free opacity dominates in the outer region. The radiation pressure dominates 
over the gas pressure in the inner region, and {\em vice versa} in the middle and outer regions. 

The temperature, gas density and disk half-thickness in each region are as follows. \\
(I) Inner region: 
\begin{equation}
 T = 3.7\times10^5\alpha^{-1/4}m^{-1/4}x^{-3/8} \ [{\rm K}], 
\end{equation}
\begin{equation}
 \rho = 6.0\ \alpha^{-1}m^{-1}\dot{m}^{-2}x^{3/2} \ [{\rm g}/{\rm cm}^{3}], 
\end{equation}
\begin{equation}
 H = 2.3\times10^{13}m\dot{m} \ [{\rm cm}]. 
\end{equation}
(II) Middle region: 
\begin{equation}
 T = 8.1\times10^6\alpha^{-1/5}m^{-1/5}\dot{m}^{2/5}x^{-9/10} \ [{\rm K}], 
\end{equation}
\begin{equation}
 \rho = 2.1\times10^{-5}\alpha^{-7/10}m^{-7/10}\dot{m}^{2/5}x^{-33/20} \ [{\rm g}/{\rm cm}^{3}], 
\end{equation}
\begin{equation}
 H = 1.9\times10^{8}\alpha^{-1/10}m^{9/10}\dot{m}^{1/5}x^{21/20} \ [{\rm cm}]. 
\end{equation}
(III) Outer region: 
\begin{equation}
 T = 1.0\times10^6\alpha^{-1/5}m^{-1/5}\dot{m}^{3/10}x^{-3/4} \ [{\rm K}], 
\end{equation}
\begin{equation}
 \rho = 1.0\times10^{-4} \alpha^{-7/10}m^{-7/10}\dot{m}^{11/20}x^{-15/8} \ [{\rm g}/{\rm cm}^{3}], 
\end{equation}
\begin{equation}
 H = 3.3\times10^{10}\alpha^{-1/10}m^{9/10}\dot{m}^{3/20}x^{9/8} \ [{\rm cm}], 
\end{equation}
\begin{equation}
 \tau_{\rm R} = 2.2\times10^3\alpha^{-4/5}m^{1/5}\dot{m}^{1/5}. 
\end{equation}
The final quantity $\tau_{\rm R}$ represents the Rosseland-mean free-free optical depth vertically 
across the disk, i.e., $\tau_{\rm R}\equiv\alpha_{\rm R}H$, which is determined self-consistently 
within the standard model.  

The radii of the boundaries between the regions I$\sim$II and II$\sim$III are given by 
\begin{equation}
 x_1=3.5\times10^2\ \alpha^{2/21}m^{2/21}\dot{m}^{16/21}, 
\end{equation}
\begin{equation}
 x_2=1.1\times10^3\dot{m}^{2/3}, 
\end{equation}
respectively. 

Based on the above equations, we discuss the physical conditions expected to appear in the putative 
standard disk in the nucleus of NGC 4261. The viscosity parameter is usually assumed to be in 
the range $\alpha=0.01\sim0.1$, and is tentatively fixed as $\alpha=0.1$ here. Other parameters 
are fixed at $m\sim5$ and $\dot{m}\sim10^{-3}$. The former value has been mentioned in the previous 
section, and the latter is set at the reference value adopted in \citet{Jon00}. These values yield 
for the radii of the boundaries, $x_1=1.5$ and $x_2=11$. This means that the major part of the accretion 
disk is in the state called the outer region, except for the innermost part ($x\lesssim 10$) where 
the standard model itself could be inaccurate owing to general relativistic effects. 

Since the occultation by the accretion disk is expected to occur at $x\gtrsim 10^3$ 
(\authorcite{Jon00}\yearcite{Jon00, Jon01}), only the outer region is of our interest here, in which 
the quantities vary like  
\begin{equation}
 T = 1.5\times10^5 x^{-3/4} \ [{\rm K}], 
\end{equation}
\begin{equation}
 \rho = 3.6\times10^{-5} x^{-15/8} \ [{\rm g}/{\rm cm}^{3}], 
\end{equation}
\begin{equation}
 n_e = 2.2\times10^{18} x^{-15/8} \ [{\rm cm}^{-3}], 
\end{equation}
where the electron number density $n_e$ is calculated by assuming full ionization for a purely 
hydrogen gas. 
At a location of $x=10^3$, for example, we have $T=8.2\times10^2$\ [K], $\rho=8.6\times10^{-12}$\ 
[g/cm$^3$], $n_e=5.2\times10^{12}$\ [cm$^{-3}$] and for the Rosseland-mean optical depth, 
$\tau_{\rm R}=4.8\times10^3$, which becomes independent of the radius in this region. 

More specifically, the frequency dependent free-free absorption coefficient is written \citep{RL79} as 
\begin{equation}
  \alpha_{\nu}^{\rm ff} = 1.8\times10^{-2}T^{-3/2}n_e^2\nu^{-2}\bar{g}_{\rm ff} \ [{\rm cm}^{-1}], 
  \label{eq:ff}
\end{equation}
where $\bar{g}_{\rm ff}$ is the velocity averaged Gaunt factor and we put $\bar{g}_{\rm ff}=1$ below 
for simplicity. The optical depth for the free-free absorption at 8.4\ GHz calculated from equations 
(\ref{eq:ff}) and (\ref{eq:tau}) is extremely large at any reasonable radius in the outer region, 
$x>x_2\simeq10$. 

Thus, the standard disk cannot provide a preferable value, $\tau\sim 1$ \citep{Jon01}, to reproduce 
the emission gap of the counterjet. Moreover, the resulting temperature is too low to maintain a 
sufficient degree of ionization in the region $x\gtrsim10^2$, which is necessary for the free-free 
processes to be important. In conclusion, the interpretation of the gap observed in the counterjet 
emission from NGC 4261 by the free-free absorption due to a standard-type accretion disk is very 
implausible. Moreover, since such a low temperature obtained above contradicts to the basic assumptions 
of the standard disks, it casts strong doubt even on the existence of a standard disk in the accretion 
flow of this object, and also of a similar type radio galaxy, at least, on the outside of the RIAF regions. 

In a similar case of NGC 6251, \citet{Sdo00} have already pointed out such difficulties as appeared 
above in interpreting the emission gap in terms of the free-free absorption by a radiation cooled disk 
of any type. Also for similar reasons, \citet{LFM03} have presented an alternative interpretation of 
this phenomenon. 

\section{Synchrotron Absorption by Resistive RIAF}

In the recent decade, a model of RIAF in a global magnetic field has proved its plausibility and 
applicability to LLAGNs (\authorcite{Kab00}\yearcite{Kab00, Kab01, Kab07}; \cite{KKY00}; \cite{YKK02}). 
Also, the direct connection of this type of accretion disks to FR\ I jets have been strongly suggested 
\citep{Kab09}. Hereafter we call it the `resistive' RIAF model, in order to distinguish it from the 
well known `viscous' RIAF model in which the presence of a global magnetic field is neglected (see, 
e.g., \cite{NMQ98}). In the former model, the resistive heating (probably of anomalous type) due to 
the electric current caused in an accretion disk plays a dominant role to dissipate the gravitational 
energy, instead of the viscous heating in the latter model. Although some type of turbulent magnetic 
fields are assumed to present in a disk also in the viscous RIAF model, it is a great advantage of 
the resistive RIAF model that this can specify the strength of the ordered magnetic field self-consistently 
within the model. In this section, we therefore try to interpret the counterjet absorptions in NGC 4261 
within the framework of the resistive RIAF model. 

The relevant quantities are given (\cite{Kab01}, \cite{YKK02}) in this model as 
\begin{equation}
 T = 9.0\times10^{11}x^{-1} \ [{\rm K}], 
\end{equation}
\begin{eqnarray}
 n_e = 9.0\times10^{10}(2n+1)^{-1/2}\Re_0^{\ 2(n+1)}x_{\rm out}^{\ -(1/2+2n)} \nonumber\\
   \times\ m^{-1}\dot{m}x^{-(1-2n)} \ [{\rm cm}^{-3}], 
\end{eqnarray}
\begin{eqnarray}
 \vert b_{\varphi}\vert = 2.4\times10^{4}(2n+1)^{-1/4}\Re_0^{\ n+1}x_{\rm out}^{\ -(1/4+n)} \nonumber\\
   \times\ m^{-1/2}\dot{m}^{1/2}x^{-(1-n)} \ [{\rm G}], 
\end{eqnarray}
where $b_{\varphi}$ denotes the azimuthal component of the magnetic field. 

Five non-dimensional parameters appeared above are as follows. The normalized mass, $m$, and the 
mass accretion rate, $\dot{m}$, are the same as before. However, since the model generally allow 
for the presence of a wind from the disk surfaces whose strength is specified by the wind parameter 
$n$ ($-1/4<n<1/2$, where negative $n$ corresponds to a down flow), the accretion rate becomes a function 
of radius. The parameter $\dot{m}$ here refers to its value at the outer edge of an accretion disk, 
$x_{\rm out}\equiv r_{\rm out}/r_{\rm S}$. In the resistive RIAF model, the radius of the outer 
edge $r_{\rm out}$ is identified as the Alfv\'en radius (which is of the same orders of magnitude 
as the Bondi radius, see \cite{Kab07}) whose value is a function of the surrounding gas temperature, 
and hence it is treated also as a parameter. The final parameter $\Re_0$ denotes the magnetic 
Reynolds number of the accretion flow, at the outer edge (the finiteness of this number corresponds 
to the name, `resistive' MHD).  

The resistive RIAF is geometrically thin in spite of its virial-like high temperature. This is because 
the magnetic pressure due to the azimuthal component, which is developed outside the disk by the disk 
rotation, compresses the plasma into a thin disk geometry (see Fig.2 of \cite{Kab07}). 
The half-opening angle of the disk is determined in terms of $\Re_0$ as $\Delta=\Re_0^{\ -(2n+1)}$. 
The inner edge radius is also specified in the model from the magnetic flux conservation as 
$x_{\rm in}=\Re_0^{-2}x_{\rm out}$. In the innermost region within this radius (i.e., $x<x_{\rm in}$), 
the basic assumptions used to derive the model becomes invalid. 
Although this region locates on the outside of the applicability of this model, this model anticipates 
in this manner the appearance of an essentially different accretion state in this innermost region. 

Here, we briefly discuss the synchrotron processes within the framework of the resistive RIAF model. 
Since the temperature in the disk ($x_{\rm in}<x<x_{\rm out}$ with $x_{\rm in}\simeq 10^2$ and 
$x_{\rm out}\simeq 10^4$) is in the range $10^8\lesssim T \lesssim10^{10}$, the electron temperature 
can be regarded as mildly relativistic. In this case, the synchrotron emissivity averaged over the
relativistic Maxwellian distribution becomes \citep{MNY96} 
\begin{equation}
 j_{\nu}d\nu = \frac{\sqrt{2}e^2n_e\omega_{\rm L}}{3c}\frac{\chi}{K_2(1/\theta_e)}
   \exp\left\{-1.89\ x_{\rm M}^{1/3}\right\}\ d\nu,
\end{equation}
where $e$ is the charge unit, $m_e$ is the electron mass, $\omega_{\rm L}\equiv eB/m_e c$ is the 
Larmor frequency, and $K_2$ denotes the modified Bessel function of order 2. Other notations are 
$\chi\equiv\omega/\omega_{\rm L}$, $\theta_e\equiv k_{\rm B}T/m_e c^2$, and $x_{\rm M}\equiv 
2\chi/3\theta_e^2$. 
Assuming a local thermodynamic equilibrium (LTE) in the accretion disk, we have for the 
synchrotron absorption coefficient in the Rayleigh-Jeans limit 
\begin{equation}
  \alpha_{\nu}^{\rm sy} = \frac{\sqrt{2}\pi e^2 c}{3k_{\rm B}}\ \frac{n_e}{\nu T}\ 
   \frac{\exp\left\{-1.89\ x_{\rm M}^{1/3}\right\}}{K_2(1/\theta_e)}. 
  \label{eq:jsy}
\end{equation}

For the discussion of NGC 4261, we set $n=0$ (i.e., the wind is neglected) for simplicity and 
$\Re_0\simeq 10$. The latter value yields the result $\Delta\simeq 0.1= 5.7^{\circ}$, which is 
consistent with the observation ($\phi/2=6.5^{\circ}$). Further, when we adopt the outer edge radius 
of $x_{\rm out}\simeq10^4$ and $\dot{m}=10^{-2}$ (this value is 10 times larger than that adopted 
by \cite{Jon00}) as suggested by our spectral fittings (see, figure \ \ref{fig:Figure2}), we obtain 
for the inner 
edge radius, $x_{\rm in}\simeq10^2$, and for the physical quantities in the disk,   
\begin{equation}
 T = 9.0\times10^{11}x^{-1} \ [{\rm K}], 
\end{equation}
\begin{equation}
 n_e = 1.8\times10^{7}x^{-1} \ [{\rm cm}^{-3}], 
\end{equation}
\begin{equation}
 \vert b_{\varphi}\vert = 3.4\times10^{2}x^{-1} \ [{\rm G}]. 
\end{equation}
The final quantity, $\vert b_{\varphi}\vert$, should be substituted for $B$ in the above synchrotron 
formulae. 

Based on this model, we first confirm that the free-free opacity is negligible in the resistive IRAF 
applied to NGC 4261. Substituting $\nu=8.4\times 10^9$ [Hz] and the expressions for $T$ and $n_e$, 
we obtain 
\begin{equation}
  \alpha_{8.4}^{\rm ff} = 9.7\times10^{-24}x^{-1/2}, \quad 
  \tau_{8.4}^{\rm ff} = 4.1\times10^{-10}x^{1/2}. 
\end{equation}
The free-free optical depth is indeed negligibly small even at $x=10^4$. 

For the observational frequency of $\nu=8.4\times10^9$ [Hz], the parameters related to the synchrotron 
process are calculated as $1/\theta_e = 6.6\times10^{-3}\ x$, $\chi\theta_e = 4.3\times10^2$, 
and $x^{1/3}_{\rm M} = 4.3\times10^{-2}\ x$. Therefore, roughly speaking, $1/\theta_e\gg1$ except in 
the region very close to the inner edge where $1/\theta_e\sim1$. In order to estimate the 
importance of the synchrotron absorption, we therefore use the asymptotic form of the modified Bessel 
function $K_2(z)\sim \sqrt{\pi/2z}\ e^{-z}$ in the limit of $z\rightarrow\infty$. In the region close 
to the inner boundary, however, the accuracy may be not so good. 

Then, the approximate form of the synchrotron absorption coefficient becomes 
\begin{equation}
  \alpha_{\nu}^{\rm sy} \simeq \frac{2\sqrt{\pi}e^2 c}{3k_{\rm B}}\ \frac{n_e}{\nu T\theta_e^{1/2}}\ 
   \exp\left\{1/\theta_e -1.89\ x_{\rm M}^{1/3}\right\}. 
\end{equation}
The existence of the exponential factor in the above formula implies that the importance of 
absorption rapidly decrease in the outer regions of the resistive RIAF where $x$ is very large.
Combining this expression with equation (\ref{eq:tau}), we can calculate the position $x$ at which 
$\tau_{8.4}^{\rm sy}\sim 1$ is realized. The result in our case gives $x = 322$. Although 
this value seems to be somewhat smaller than the observational value of $x\simeq10^3$ \citep{Jon00}, 
we cannot say too much because the above treatment is too crude, especially, near the disk's inner 
boundary ($x_{\rm in}\sim 100$, see the next section) which is close to the estimated radius of 
$\tau_{8.4}^{\rm sy}\sim 1$.

Nevertheless, we can say from the above discussion that the resistive RIAF model predicts the 
existence of rather narrow region near the disk's inner edge, in which the synchrotron absorption becomes 
appreciable. Therefore, this region can contribute to the formation of such a narrow gap as observed 
in the counterjet emission from NGC 4261 (and also from similar objects). 
This point will be further discussed in the next section in relation to detailed numerical 
calculations of the emission spectrum.
Within the inner edge radius, the accretion flow is expected to form a condensed cool disk owing 
to efficient synchrotron cooling. Such an inner disk might also contribute to the absorption 
of counterjet emission. However, detailed discussions have to await the appearance of a concrete inner 
disk model. 

\section{Radiative Processes in Resistive RIAF}

Although a crude estimation of synchrotron processes has indicated in the previous section 
a very good possibility of interpreting the gap seen in the counterjet emission of NGC\ 4261 as 
synchrotron absorption by a foreground resistive RIAF disk, further detailed confirmation of this 
possibility is given in this section based on numerical treatments of the relevant processes within 
the framework of the resistive RIAF model.

\subsection{Reproduction of global SED}

\begin{figure}
 \begin{center}
  \FigureFile(88mm, 80mm){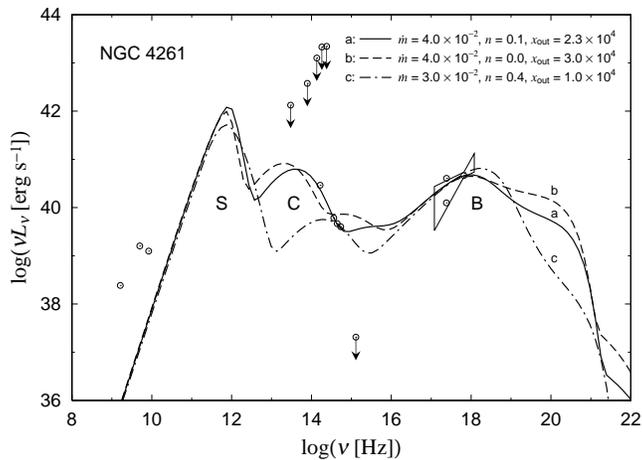}
 \end{center}
 \caption{Fittings to the observed spectrum of NGC 4261. The model curves are calculated 
 based on our resistive RIAF model. The values of varied parameters are cited in the figure 
 and those of fixed parameters are  $m=5$ and $\Re_0=10$ (or $\Delta=6^{\circ}$).}
\label{fig:Figure2}
\end{figure}

In this subsection, we want to prove persuasive abilities of the resistive RIAF model by reproducing 
the observational wide range SED.
In figure \ref{fig:Figure2}, we plot luminosity $\nu L_{\nu}$ against frequency $\nu$. 
Each curve there represents only the radiation from an accretion disk of the RIAF type which 
extends from the outer edge radius, $r_{\rm out}=x_{\rm out}r_{\rm S}$ (roughly corresponds to the 
Bondi radius), down to the inner edge radius, $r_{\rm in}=x_{\rm in}r_{\rm S}$. In actual situations, 
however, there may be other contributions, e.g., from bipolar jets and from an inner cool disk. 
The observational data are adopted from \citet{Ho99} and \citet{Bal06}. Since the radiation in the 
radio wave band is very likely to come mainly from the jet component, we try to fit our calculated 
curves to the data points in the two frequency ranges, that from near-infrared (NIR) to 
infrared (IR) and that in X-rays. 

The resistive RIAF model has 5 parameters to specify. They are mass of the central black hole $m$, outer 
edge radius of the disk $x_{\rm out}$, mass accretion rate (at the outer edge) $\dot{m}$, magnetic 
Reynolds number (at the outer edge) $\Re_0$, and wind parameter $n$. As already described, we can roughly 
fix two parameters as $m=5$ and $\Re_0=10$ (or $\Delta=\phi/2\simeq 0.1=5.7^{\circ}$) based on observations. 
The values of the remaining parameters are varied and are cited in figure \ref{fig:Figure2} 
for each curve shown there. As seen from these curves, the disk radiation consists 
mainly of three peaks in different frequency ranges. They are labeled as S, B, and C for convenience, 
whose emission mechanisms are attributed, respectively, to thermal synchrotron emission, thermal 
bremsstrahlung, and their inverse Compton scatterings. 

The method for calculating these contributions are given in \citet{KKY00}, except some improvements added 
later. The detailed expressions of the Gaunt factor in extremely relativistic bremsstrahlungs have been 
properly taken into account in \citet{YKK02}, and the synchrotron emissivity for mildly relativistic 
cases (equation (23) in this paper) is also taken into account in this paper in addition to that for 
extremely relativistic cases (equation (21) in \cite{KKY00}). 

In figure \ref{fig:Figure2}, curve {\it a} represents our best fit case, in which the curve fits very good 
not only to the X-ray luminosity and slope there but also to 4 points in IR and NIR range.  
The presence of inverted spectrum on the lower side of the synchrotron peak frequency clearly indicates 
that the radiation becomes opaque in this frequency region. Since the synchrotron emission becomes 
strong only where magnetic field is strong and temperature is high, it comes mainly from inner parts 
of the disk. The location of this inverted slope in the diagram is determined predominantly by the mass 
parameter $m$ only. Therefore, we can say that the excess seen in the radio frequency range should be 
attributed to the components other than the accretion disk, most probably, to bipolar jet (\cite{Yun07}), 
unless the estimated black hole mass turns out to be larger by a few orders of magnitude. 

A plateau like peak B is caused by thermal bremsstrahlung process, for which the low frequency side 
comes from outer parts of the accretion disk while the higher side comes from its innermost part. 
Peak C on the right hand side of peak S is due to the synchrotron photons once up-scattered by 
relativistic thermal electrons near the inner edge. We can also see on the higher side of C a slight 
peak caused by twice up-scattered photons. Such a peak is more evident on curve {\it b}. The effects of 
further multiple scatterings are neglected. Similarly to the case of synchrotron photons, the effects 
of inverse Compton scattering on the bremsstrahlung photons can be recognized in the frequency region 
above $\nu\sim 10^{21}$ Hz. 

Beside our best-fit curve {\it a}, we show other curves {\it b} and {\it c} for comparison. Although 
the latter two curves also show good fits to the X-ray data, they have only poorer fits than the 
former in the IR range. Indeed, both curves do not go through the NIR point, and further, curve {\it b} 
goes through only one IR point out of three. Since curve {\it b} is a restricted fit within no wind 
condition ($n=0$), the results strongly suggests that the presence of a disk wind in this object even 
if it may be week ($n\simeq 0.1$). Curve {\it c} is also a restricted fitting in which the outer edge 
radius is arbitrarily fixed at $x_{\rm out}=10^4$. Although the difference in the values of $x_{\rm out}$ 
from the best-fit case is only a factor of 2, we need an extremely strong wind ($n=0.4$) in this case. 
This may imply that the disk's outer edge radius can be determined fairly well by this kind of 
spectral fittings. 

It is instructive to note the change in the shape of peak B according to an increase in the wind 
parameter $n$. As seen in the figure, it results in a decrease of the higher-frequency side shoulder 
of B.
This is indeed an expected result, because as the disk wind becomes strong the matter that reaches deep 
down the accretion disk decreases and hence radiation produced there also decreases. 
Another point to mention is that for NGC 4261 we cannot find any explicit indication of 
a cool disk, which may be seen as an enhancement of the data points in ultra-violet 
(UV) range.

Thus, we have shown above that the radiation from a resistive RIAF disk can explain the main 
features of the observed spectrum of NGC 4261, and the parameters adopted in the previous section are 
very plausible ones. 

\subsection{Resistive RIAF as an absorber}

\begin{figure}
 \begin{center}
  \FigureFile(88mm, 80mm){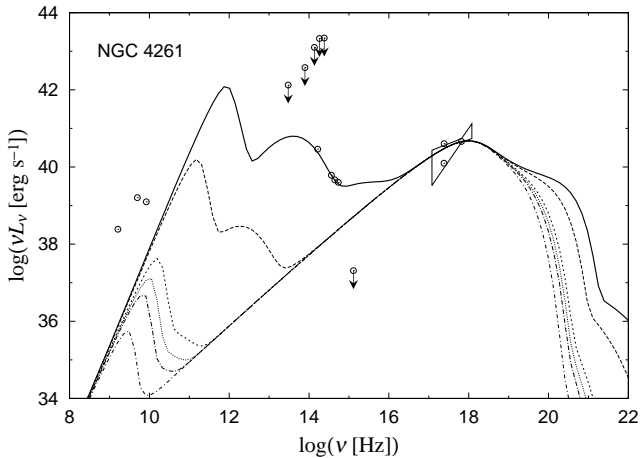}
 \end{center}
 \caption{Contributions to SED from different partial disks whose inner edges are artificially 
 truncated at several radii. The full curve show the original SED which is the best fit case in 
 figure 2, whose inner edge radius is written as $r_{\rm in}$. 
 The dotted curves correspond from above to the truncations at $2r_{\rm in}$, $5r_{\rm in}$, $6r_{\rm in}$, 
 $7r_{\rm in}$ and  $10r_{\rm in}$, respectively. }
\label{fig:Figure3}
\end{figure}

Since a resistive RIAF is in LTE, a portion of good emitter is also a good absorber of the same 
radiation field. When it is placed in front of the counterjet, the jet emission is absorbed effectively 
by the part of the disk which can produce the corresponding emission. 
Therefore, the problem here is what part of the RIAF disk can contribute to the emission at 8.4\ GHz. 
Since the synchrotron process is the dominant mechanism to produce this radio emission in a resistive 
RIAF, it works effectively in the place where both temperature and magnetic field are large. 
This corresponds to the innermost part of the disk. In order to evaluate the most efficient part of 
the accretion disk to produce the 8.4\ GHz radio emission, specifically, we show in figure 
\ref{fig:Figure3} different contributions to the total SED from several different partial disks. 
The dashed curves represent from the top the contributions from the disks whose inner edges are 
artificially truncated at $2r_{\rm in}$, $5r_{\rm in}$, $6r_{\rm in}$, $7r_{\rm in}$, $10r_{\rm in}$, 
respectively, while the full curve does the original SED. 

Therefore the difference, for example, between the uppermost curve (the full curve) and the second 
curve (the uppermost dashed curve) represents the emission produced in the annular part of the disk with 
its radius in between $r_{\rm in}$ and $2r_{\rm in}$. It is evident from this figure that this innermost 
part of the disk contributes to the highest frequency parts of the synchrotron and bremsstrahlung emissions. 
The part which contributes mainly to 8.4GHz turns out from this figure to be the annulus between 
$6r_{\rm in}$ and $10r_{\rm in}$. Therefore, this part also acts most effectively in absorbing the radio 
of this frequency. Since the best fit parameters of the model SED indicate that the inner edge radius is 
$x_{\rm in}=2.3\times 10^2$, the absorbing annulus corresponds to 
$1.4\times10^3 r_{\rm S} < r < 2.3\times10^3 r_{\rm S}$. 
This size seems to be consistent with the observed distance from the core to the absorption gap 
(a few $\times10^3 r_{\rm S}$, see \cite{Jon01}). 

If we define the spectral index of radiation as $\beta \equiv d \ln F_{\nu}/d \ln\nu$, it is related 
to the slope of the SEDs adopted in the present paper as $d \ln \{\nu F_{\nu}\}/d \ln\nu = 1 + \beta 
\equiv \gamma$. 
The part of the disk which produces the radiation of a positive spectral index ($\beta>0$ or $\gamma>1$) 
is optically thick at that frequency and emitted radiation is self-absorbed. 
Therefore, we can read from figure 3 that the innermost part of the disk ($r_{\rm in}< r < 7r_{\rm in}$) 
is evidently opaque to the radiation near 8.4\ GHz although its contribution to the radiation spectrum 
at these frequencies is rather small. 

Figure 9 of \citet{Jon01} show that the spectral index of the counterjet emission becomes very large 
(probably, $\beta>2.5$) near the peak absorption, suggesting that it exceeds the value of self-absorbed 
synchrotron emission. Similar steep indices are observed also at the inner edge of the jet emission 
in NGC\ 1052 and interpreted as due to the free-free absorption by a circumnuclear torus (e.g., 
\cite{Kam05, SSS08}). 
Although these authors insist that such a steep index favors the interpretation in terms of free-free 
absorption by a foreground plasma, the same logic as theirs also apply to synchrotron absorption. 
It is a result of selective absorption of lower frequency parts caused by obscuring matter. 

Owing to such a process, the incoming intensity $I_{\rm in}(\nu)$ attenuates like 
$I_{\rm in}\exp\{-\tau(\nu)\}$. Here $\tau(\nu)$ is the optical depth across the absorber, which is 
written as $\tau^{\rm sy}(\nu)\simeq(1/2)\alpha^{\rm sy}_\nu L$ in our case. As seen from equation 
(\ref{eq:jsy}), the frequency dependence of the synchrotron absorption coefficient is 
\begin{equation}
 \alpha^{\rm sy}_\nu \propto \nu^{-1}\exp(-\tau_0 \nu^{1/3}) 
\end{equation}
with $\tau_0$ being a constant, since $x_{\rm M}\propto\nu$. 
Thus, the attenuation becomes rapidly small as the frequency increases.

\section{Summary and Conclusion}

First, we have examined the plausibility of the assertion that the very narrow gap observed at radio 
frequencies in the counterjet emission of NGC 4261 is caused by the free-free absorption of a radiatively 
cooled accretion disk around its central black hole. 

If the radiatively cooled disk is assumed to be optically thin, then the self-consistent model (SLE 
model) predicts that such a disk cannot be persistent, since it is known to be thermally unstable. 
In addition, such a disk is not geometrically thin and the optical depth is extremely small. 
Both these facts are unfavorable to explain the observations. 
On the other hand, if the disk is assumed to be an optically thick disk of the standard type, then the 
optical depth becomes too large and, further, the temperature in the disk turns out to be too small 
to maintain a sufficient ionization state, except in the region very close to the center. 

Therefore, we may conclude that the emission gap is very unlikely to be caused by the free-free absorption 
of a radiatively cooled accretion disk of any type. Also, the low temperatures derived from the standard 
disk model strongly suggests the implausibility of realizing such disks, at least, in the outer accretion 
regions of NGC 4261 specifically, and also of similar type radio galaxies and of LLAGNs in general, 
all of which are accreting with very small fraction of each Eddington rate. 

Next, we have examined the possibility of explaining the emission gap by a RIAF in a global magnetic 
field (i.e., resistive RIAF), and found that the model can actually explain both the appearance 
of the gap and the distance of peak absorption from the core, in terms of the synchrotron absorption 
caused by an inner part of the accretion disk. The model can also reproduce the observed global SED very 
well, except in the radio frequency range where the contribution from the jet is expected to be large.

Thus, we may conclude that the standard-type accretion disks, at least in its original form, do not 
exist anywhere in LERGs and nearby LLAGNs. Instead, the state of accretion flows in such objects 
are very likely to be described by the resistive RIAF model. 

The resistive RIAF model predicts that a hot RIAF state is directly connected to a surrounding hot 
gas phase at around the Bondi radius from the center. In the case of NGC\ 4261, this radius is estimated 
to be $\sim 2.3\times 10^4 r_{\rm S}$ from our spectral fittings, and is somewhat smaller than the inner 
radius of the geometrically-thin H\ I disk found by VLBI observations \citep{vLan00}. 
Although the H\ I disk seems to be connected to the outer HST dust torus \citep{Jaf93}, its relation 
to the accretion disks discussed in the present paper is still unclear. 

The presence of such a circumnuclear tours is inferred also in NGC\ 1052 from radio observations 
\citep{Kam05, SSS08}, and is expected to supply matter to the accretion disk enclosed within it.  
The central depression of the jet emission in this object and the steepening of its spectrum are 
interpreted as due to free-free absorption by this torus.

The resistive RIAF model also predicts that such an RIAF state cannot extends down to the last stable 
orbit around the central black hole, but terminates at the inner edge located at about $100\ r_{\rm S}$, 
where the effective magnetic Reynolds number becomes unity. 
Within this radius, the disk would become optically thick to synchrotron absorption up to 
$\sim10^{12}$ Hz and expected to cool down to a temperature much lower than the virial one. Such a cool 
inner disk may resemble the standard disk, but, as already pointed out by some authors (e.g., 
\cite{KnB07}), its radiation spectrum would be largely modified on account of the energy extraction 
caused by fairly strong vertical magnetic fields concentrated in this inner region. The detailed 
understanding of such processes, however, belongs to an attractive future work. 

We would like to thank Seiji Kameno, the referee of this paper, for his several comments which 
were very useful to improve the manuscript.
This work is partly supported by the Gant-in Aid for Scientific Research (C; 20540233, K.~W.) 
from the Japanese Ministry of Education, Culture Sports, Science and Technology (MEXT).

\end{document}